\documentclass[reprint,superscriptaddress,amsmath,amssymb,aps,prx]{revtex4-1}

\usepackage{amsmath}
\usepackage{amsthm}
\usepackage[colorlinks=true,urlcolor=blue,citecolor=blue,linkcolor=blue]{hyperref}
\usepackage[shortlabels]{enumitem}
\usepackage{amsfonts}
\usepackage{graphicx}
\usepackage{bbold}
\usepackage{stmaryrd}
\usepackage{color}
\usepackage{hyperref}
\usepackage{verbatim}
\usepackage{cases}
\usepackage{amsfonts}
\usepackage{amssymb}
\usepackage[]{mathrsfs}
\usepackage[table]{xcolor}
\usepackage{epsfig}
\usepackage{bm}
\usepackage{setspace}
\usepackage{enumerate}
\usepackage{dcolumn}
\usepackage{bm}
\usepackage{enumitem}

\usepackage{graphicx}
\usepackage{dcolumn}
\usepackage{bm}
\usepackage{color}
\usepackage{amssymb}
\usepackage{lipsum}
\usepackage{amsmath}
\usepackage{graphicx}
\usepackage{array}
\usepackage{multirow}
\usepackage{booktabs}

\usepackage[normalem]{ulem}

\newcommand{\bra}[1]{\left< #1 \right\vert}
\newcommand{\ket}[1]{\left\vert #1 \right>}
\newcommand{\pare}[1]{\left( #1 \right)}
\newcommand{\abs}[1]{\left\vert #1 \right\vert}
\newcommand{\cor}[1]{\left[ #1 \right]}

\newcommand{\llav}[1]{\left\lbrace #1 \right\rbrace}

\begin{document}

\preprint{APS/123-QED}

\title{Reconfigurable Network for Quantum Transport Simulation}

\author{Mario A. Quiroz-Ju\'{a}rez}
\affiliation{Instituto de Ciencias Nucleares, Universidad Nacional Aut\'onoma de
M\'exico, Apartado Postal 70-543, 04510 Cd. Mx., M\'exico}

\author{Chenglong You}
\affiliation{Department of Physics and Astronomy, Louisiana State University, Baton Rouge, Louisiana 70803, USA}

\author{Javier Carrillo-Mart\'{i}nez}
\affiliation{Instituto de Ciencias Nucleares, Universidad Nacional Aut\'onoma de
M\'exico, Apartado Postal 70-543, 04510 Cd. Mx., M\'exico}

\author{Diego Montiel-\'{A}lvarez}
\affiliation{Instituto de Ciencias Nucleares, Universidad Nacional Aut\'onoma de
M\'exico, Apartado Postal 70-543, 04510 Cd. Mx., M\'exico}



\author{Jos\'e L. Arag\'on}
\affiliation{Centro de F\'isica Aplicada y Tecnolog\'ia Avanzada, Universidad Nacional Aut\'onoma de
M\'exico, Boulevard Juriquilla 3001, 76230 Quer\'etaro, Juriquilla, M\'exico}

\author{Omar S. Maga\~{n}a-Loaiza}
\affiliation{Department of Physics and Astronomy, Louisiana State University, Baton Rouge, Louisiana 70803, USA}

\author{Roberto de J. Le\'on-Montiel}
\affiliation{Instituto de Ciencias Nucleares, Universidad Nacional Aut\'onoma de M\'exico, Apartado Postal 70-543, 04510 Cd. Mx., M\'exico}

\date{\today}

\begin{abstract}

In 1981, Richard Feynman discussed the possibility of performing quantum mechanical simulations of nature. Ever since, there has been an enormous interest in using quantum mechanical systems, known as quantum simulators, to mimic specific physical systems. Hitherto, these controllable systems have been implemented on different platforms that rely on trapped atoms, superconducting circuits and photonic arrays. Unfortunately, these platforms do not seem to satisfy, at once, all desirable features of an universal simulator, namely long-lived coherence, full control of system parameters, low losses, and scalability. Here, we overcome these challenges and demonstrate robust simulation of quantum transport phenomena using a state-of-art reconfigurable electronic network. To test the robustness and precise control of our platform, we explore the ballistic propagation of a single-excitation wavefunction in an ordered lattice, and its localization due to disorder. We implement the Su-Schrieffer-Heeger model to directly observe the emergence of topologically-protected one-dimensional edge states. Furthermore, we present the realization of the so-called perfect transport protocol, a key milestone for the development of scalable quantum computing and communication. Finally, we show the first simulation of the exciton dynamics in the B800 ring of the purple bacteria LH2 complex. The high fidelity of our simulations together with the low decoherence of our device make it a robust, versatile and promising platform for the simulation of quantum transport phenomena.

\end{abstract}

\pacs{Valid PACS appear here}
\maketitle


Understanding the limits of controllability of quantum and classical transport has long been considered a topic of great relevance in physics, chemistry, and biology \cite{hanggi2009,lambert2012,cao2020}. In particular, the study of novel materials that exhibit excitation-energy transfer pathways, that are different from those available in nature, has recently attracted a great deal of attention. Indeed, the control of transport phenomena at the nanoscale has shown an enormous potential for the development of new light-harvesting technologies for solar energy conversion \cite{photo_book}, enhanced sensing \cite{hodaei_2017,pirandola2018,chen2020}, and even for the design of electronic and photonic circuits capable of performing complex tasks with high efficiency \cite{aspuru2012,lee2018,wang2020,elshaari2020}. In this regard, quantum random walks have emerged as useful tools for the experimental simulation of non-trivial transport phenomena. In general, quantum networks have been implemented on different platforms, such as optical cavities \cite{caruso2011,viciani2015,beaudoin2017}, trapped ions \cite{zahringer2010,blatt2012,trautmann2018}, ultracold atomic lattices \cite{lewenstein2007,bloch2012,preiss2015,dadras2018}, superconducting circuits \cite{peropadre2016,chin2018,yan2019,kjaergaard2020}, and integrated photonics \cite{schreiber2011,sansoni2012,crespi2013,rechtsman2013,caruso2016,armando2018,harris2017,alan2019,magana2019multiphoton,you2020multiparticle}.

Unfortunately, these platforms do not seem to satisfy, at once, all desirable features of a universal simulator, namely full control of the system's parameters, low losses, and scalability. In this work, we demonstrate robust simulation of quantum transport using a state-of-art reconfigurable electronic network. This is managed by constructing a unique mapping that allows us to establish a direct connection between the probability amplitudes of a quantum tight-binding system and the voltages of coupled electrical-oscillator networks. Our platform, which comprises ten fully reconfigurable RLC oscillators, is implemented by means of operational amplifiers and passive linear electrical components. This let us operate, as many of the aforementioned platforms, within the single-excitation Hilbert subspace, i.e., the space that describes the dynamics of a single particle in a tight-binding quantum network \cite{roberto-LPL}.

To test the versatility and precision of our platform, we have implemented different quantum transport protocols that demand specific site-frequencies and coupling conditions. In particular, we have explored the ballistic propagation of a single-excitation wavefunction in an ordered lattice and its localization due to stochastically-varying couplings (static disorder), the so-called Anderson localization \cite{anderson1958}. We have implemented the Su-Schrieffer-Heeger (SSH) model \cite{ssh1,ssh2}, where the proper use of alternating-coupling values, and fixed site-energies, allows us to directly observe the emergence of one-dimensional edge states. Because of its relevance for scalable quantum computing and communication \cite{nielsen_book,liang2005}, we have implemented the protocol known as perfect transport \cite{chris2004,armando2013}, which makes use of a linear chain of qubits (or sites) where the couplings between them follow a precise square-root rule to coherently transfer quantum states.
Finally, we have tested our platform capabilities for mimicking the transport behavior of photosynthetic light-harvesting complexes by implementing the first simulation of the exciton dynamics in the B800 ring of the purple bacteria LH2 complex \cite{cheng2006}.


The dynamics of a single excitation in a system comprising $N$ coupled quantum oscillators is described by the Schrodinger equation $i\partial_t\ket{\psi\pare{t}} = \hat{H}\ket{\psi\pare{t}}$, where the Hamiltonian is given by
\begin{equation}\label{Eq:Hamiltonian}
\hat{H}=\sum_{n=1}^{N}\varepsilon_n\ket{n}\bra{n} + \sum_{n\neq m}^{N}J_{nm}\ket{n}\bra{m},
\end{equation}
with $\ket{n}$ denoting the energy density associated to the $n$th oscillator. The $n$th-site energies and the coupling between sites $n$ and $m$ are given by $\varepsilon_n$ and $J_{nm}$, respectively. Then, by expanding the time-dependent wavefunction in the site basis, i.e. $\ket{\psi\pare{t}} = \sum_{n}c_{n}\pare{t}\ket{n}$, it is straightforward to find that the Schrodinger equation leads to a set of first-order coupled differential equations of the form $i\partial_{t}c_{n} = \varepsilon_{n}c_{n} + \sum_{n\neq m}^{N}J_{nm}c_{m}$. In the weak-coupling limit ($J_{nm}\ll \varepsilon_{n}$), the time-derivative of this equation becomes \cite{briggs2011,roberto2013}
\begin{equation}\label{Eq:second_dev_Q}
\frac{d^{2}c_{n}}{dt^{2}} = -\varepsilon_{n}^{2}c_{n} - \varepsilon_{n}\sum_{n\neq m}^{N}2J_{nm}c_{m}.
\end{equation}
As we will show below, the importance of this expression resides in the fact that it allows us to establish a direct connection between the probability amplitudes $c_{n}$ of a quantum system and the voltages $V_{n}$ in an electrical-oscillator network. To do so, let us consider an array of $N$ inductively-coupled RLC oscillators (see Supplementary Materials for details), where R, L and C stand for resistor, inductor and capacitor, respectively. We can use the Kirchhoff laws to find that the equations of motion for the voltages $V_{n}\pare{t}$ across the capacitors $C_{n}$ are given by
\begin{equation}\label{Eq:second_dev_C}
\begin{split}
\frac{d^{2}V_{n}}{dt^{2}} = \frac{1}{C_{n}} & \left[- \frac{1}{R_{n}}\frac{dV_{n}}{dt} - \frac{V_{n}}{L_{n}} - \sum_{j=n+1}^{N}\frac{V_{n}-V_{j}}{L_{nj}}\right. \\
 & \left. \hspace{2mm} + \sum_{j=1}^{j<n}\frac{V_{j} - V_{n}}{L_{jn}} \right],
\end{split}
\end{equation}
where $L_{nj}$ stands for the inductor that couples the $n$th and $j$th oscillators. Remarkably, by writing Eq. (\ref{Eq:second_dev_C}) in the non-dissipative limit, i.e. when $R\rightarrow \infty$, one can find that it is mathematically equivalent to Eq. (\ref{Eq:second_dev_Q}) with
\begin{equation}\label{Eq:Energies}
\varepsilon_{n}^{2} = \frac{1}{C_{n}}\pare{\frac{1}{L_{n}} + \sum_{m\neq n}^{N}\frac{1}{L_{nm}}}, \hspace{2mm} J_{nm} = -\frac{1}{2\varepsilon_{n}L_{nm}C_{n}}.
\end{equation}
This mapping among probability amplitudes, $c_{n}\pare{t}$, and voltages, $V_{n}\pare{t}$, is then completed by adding a non-Hermitian term to the Hamiltonian (\ref{Eq:Hamiltonian}), which accounts for the parasitic losses that are present in its experimental implementation. It is worth mentioning that this non-Hermitian term is determined by analyzing the time-dependent energy in the quantum and electronic models. While the total energy in the quantum system is given by $Q_{\text{q}}(t)=\sum_{n}\left|c_n\right|^2$, the energy stored in and across the ten coupled oscillators is obtained by writing $Q_{\text{cl}}(t)=\frac{1}{2}\sum_{m \neq n} C_nV_n^2+L_nI_n^2+L_{nm}I_{nm}^2$, where $I_{n}$ and $I_{nm}$ stand for the currents passing through the oscillator and coupling inductors, respectively. By tracking time-traces for both energies, one can find that the energy decay rates keep a quantitative agreement if the term $\hat{H}_{\text{loss}} = -\frac{i}{2}\sum_{n}\Gamma_{n}\ket{n}\bra{n}$, with $\Gamma_{n} = 1/\pare{R_{n}C_{n}}$ describing the rate at which energy is dissipated, is included in the Hamiltonian (\ref{Eq:Hamiltonian}), see Supplementary Materials for details.



Our current version of the electronic platform comprises ten fully reconfigurable RLC oscillators, where site-frequencies and couplings can be independently selected from a broad range of possible values (see Supplementary Materials and Refs. \cite{roberto2015,alan2017,roberto2018} for details). This allows us to explore different quantum transport protocols, including Anderson localization, the emergence of edge states in the SSH model, the coherent transfer of a quantum state, and the simulation of excitonic energy transport in photosynthetic light-harvesting complexes.

\begin{figure}[b!]
\centering
\includegraphics[width=8.25cm]{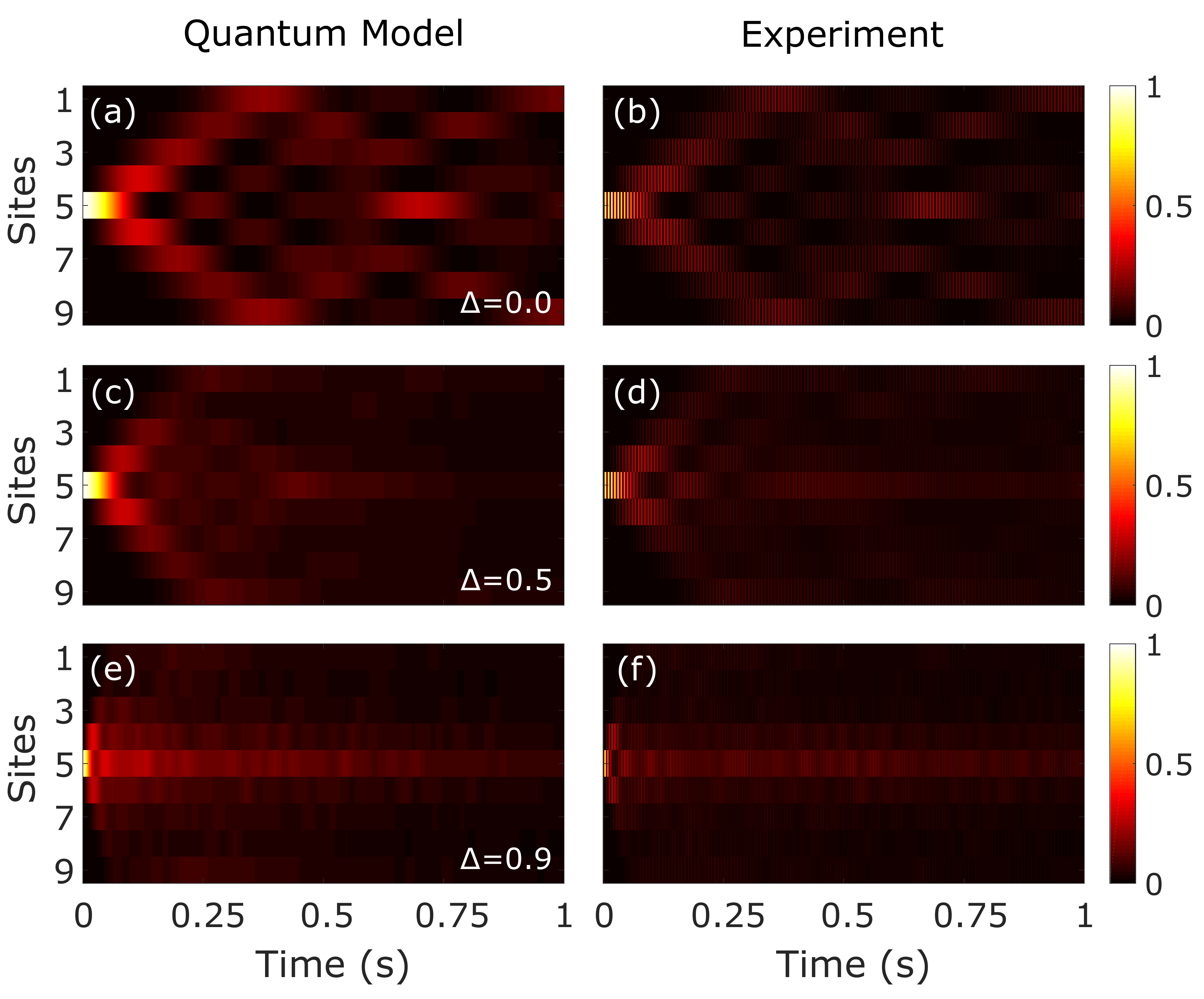}
\caption{Time evolution of an excitation (voltage signal) initialized in the central site of a one-dimensional network comprising nine nearest-neighbor-coupled oscillators. The rows (from top to bottom) depict the evolution for increasingly larger degree of disorder: (a-b) $\Delta = 0$, (c-d) $\Delta = 0.5$, and (e-f) $\Delta = 0.9$. The results presented in all panels correspond to the average of 50 different disordered-array time evolutions.}
\label{Fig:Anderson_results}
\end{figure}

\emph{Anderson Localization.-} The localization of a particle's wavefunction in disordered lattices is one of the most fascinating effects in physics \cite{anderson1958}. This fundamental phenomenon, known as Anderson localization, arises from the interference of multiple scattering effects. In this scenario, the wavefunction of a propagating particle in a lattice is affected by static disorder, introduced in either the lattice-site energies (diagonal disorder) or in the coupling among them (off-diagonal disorder) \cite{armando2011}. We have implemented the Anderson localization protocol by making use of $N=9$ out of the ten available oscillators in our electronic platform. The oscillators are arranged in a one-dimensional nearest-neighbor-coupled lattice described by the Hamiltonian
\begin{eqnarray}
\hat{H}_{\text{AL}} & = & \sum_{n=1}^{N}\pare{\varepsilon_n-i\Gamma_{n}}\ket{n}\bra{n} + \sum_{n = 1}^{N-1}J_{n,n+1}\ket{n}\bra{n+1} \nonumber \\
& + & \sum_{n = 1}^{N-1}J_{n+1,n}\ket{n+1}\bra{n}.
\end{eqnarray}
All site frequencies $\varepsilon_n$ and losses $\Gamma_{n}$ are described by the values presented in the first row of Table I in the Supplementary Materials. The static disorder is introduced through the coupling between sites by randomly selecting the value of each coupling inductor from a uniform distribution. This is described by $\cor{L_{x}\pare{1 - \Delta}, L_{x}\pare{1 + \Delta}}$ with $L_{x}=96.05$ mH and $\Delta = 0,0.5,0.9$ indicating the degree of the lattice disorder. Figure \ref{Fig:Anderson_results} shows the time evolution of an excitation (voltage signal) initialized in the central site of the one-dimensional lattice. The first column shows the quantum-mechanically-predicted population $\abs{c_{n}}^2$ evolution, whereas the second column shows our experimentally-obtained squared-voltage-signal $\abs{V_{n}}^2$ evolution. Notice that, as one might expect, ballistic propagation of the excitation is observed when disorder is absent ($\Delta = 0$), see Figs. \ref{Fig:Anderson_results}(a-b); while for strong disorder ($\Delta=0.9$) the excitation gets localized in the central site of the lattice, as depicted in Figs. \ref{Fig:Anderson_results}(e-f). It is important to remark that given the stochastic nature of Anderson localization, the results shown in each panel of Fig. \ref{Fig:Anderson_results} correspond to the average of 50 different disordered-array time evolutions.

\begin{figure}[t!]
\centering
\includegraphics[width=8.0cm]{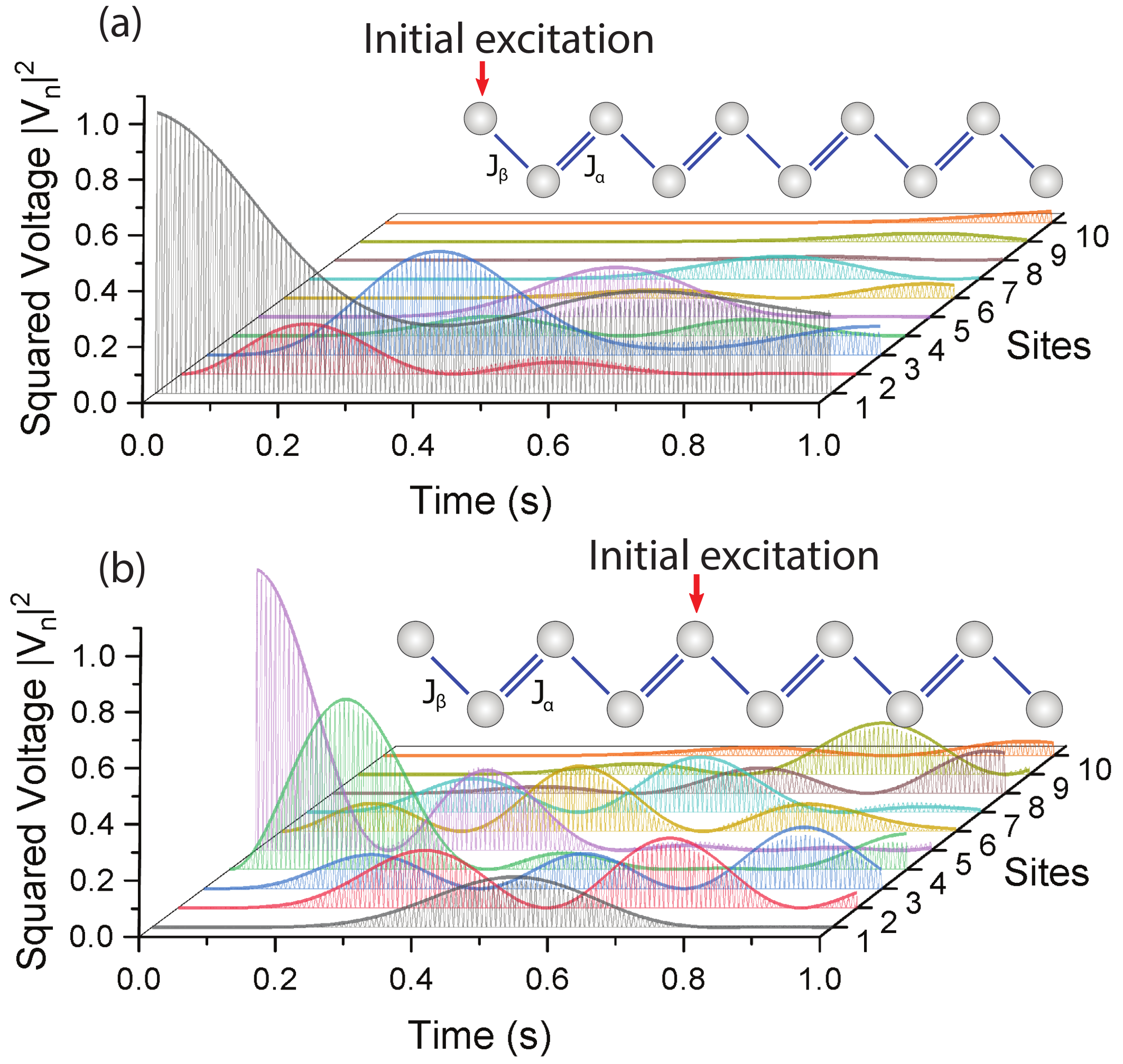}
\caption{Time evolution of a voltage signal (excitation) initialized in (a) the edge and (b) the bulk of a ten-oscillator Su-Schrieffer-Heeger (SSH) chain. The fast oscillating signals correspond to the experimentally-measured squared voltages $\abs{V_{n}}^2$, whereas the slowly-varying envelope (solid lines) shows the theoretically-predicted behavior of the quantum populations $\abs{c_{n}}^2$. The insets show the lattice structure, as well as the initial excitation conditions in each case.}
\label{Fig:SSH_results}
\end{figure}

\emph{The Su-Schrieffer-Heeger (SSH) Model.-} One of the simplest models to study non-trivial topology phenomena, such as the emergence of topologically-protected edge states, is the Su-Schrieffer-Heeger (SSH) model \cite{ssh1,ssh2}. The SSH model describes the hopping of a spinless fermion on a one-dimensional lattice with staggered hopping amplitudes, as shown in the insets of Fig. \ref{Fig:SSH_results}. The chain consists of $N$ unit cells, each of which hosts two sites, one on sublattice $A$, and one on sublattice $B$. We neglect interactions between electrons, consequently the dynamics of each electron is described by a single-excitation Hamiltonian of the form \cite{asboth2016}
\begin{eqnarray}\label{Eq:SSH}
\hat{H}_{\text{SSH}} &=& \sum_{n=1}^{N}\cor{\pare{\varepsilon_n-i\Gamma_{n}}\ket{n}\bra{n} + J_{\alpha}\pare{\ket{n,B}\bra{n,A} + \text{H.c.}}} \nonumber \\
&+& J_{\beta}\sum_{n = 1}^{N-1}\pare{\ket{n+1,A}\bra{n,B} + \text{H.c.}},
\end{eqnarray}
where the states of the chain are described by $\ket{n,A}$ and $\ket{n,B}$, with the electron's unit cell represented by $n\in\llav{1,2,...,N}$, and H.c. stands for the Hermitian conjugate.

Arguably, the most important feature of the SSH model is the emergence of topologically protected edge modes at the end of the chain, when the intracell coupling $J_{\alpha}$ exceeds the intercell coupling $J_{\beta}$ \cite{Liu2019}. We have experimentally produced these states by making use of all ten sites in our electronic platform. The parameters used for implementing the Hamiltonian in Eq. (\ref{Eq:SSH}) are described in the second row of Table I in the Supplementary Materials. Note that the coupling inductors satisfy the condition $J_{\alpha} = 2J_{\beta}$. Figure \ref{Fig:SSH_results} shows the time evolution of an excitation (voltage signal) initialized in (a) the edge and (b) the bulk of the chain. Note that the fast oscillating signals correspond to the experimentally-measured squared voltages $\abs{V_{n}}^2$. The slowly-varying envelope (solid lines) shows the theoretically-predicted behavior of the quantum populations $\abs{c_{n}}^2$. These results demonstrate two important facts: (1) the quantum probabilities follow the same dynamics as the envelope of the squared voltage signals, and (2) the small frequencies used in our device ($\sim$ $1.5$ kHz) allow for a rather simple extraction of the amplitude and phase of the signals. In general, this is a cumbersome task in experiments working at optical (or higher) frequencies. Finally, note from Fig. \ref{Fig:SSH_results} that the relation $J_{\alpha}=2J_{\beta}$ creates a condition in which any excitation initialized in the edge will tend to stay there for a longer time than when injecting energy in any site of the bulk. This is precisely the result of the topological edge protection \cite{ssh1,ssh2}. It is important to remark that, as in other topological-insulator examples, this energy-localization effect becomes stronger as the system's size is increased \cite{bookLuo,pablo2020}.

\emph{Coherent Transfer of States.-} A key milestone for the development of scalable quantum computing and communication is the coherent transfer of states among numerous sites in an extended network. Remarkably, it has been shown that if coherence is maintained across many sites, the transfer of quantum states can be obtained with extremely high efficiency \cite{nielsen_book,bose2003,bose2007,kay2010}. Indeed, this so-called \emph{perfect state transfer} can be observed by engineering a qubit-chain described by a Hamiltonian of the form \cite{chris2004,armando2013}
\begin{equation}\label{Eq:PT}
\begin{split}
\hat{H}_{\text{CT}} & = \sum_{n=1}^{N}\pare{\varepsilon_n-i\Gamma_{n}}\ket{n}\bra{n} + \sum_{n = 1}^{N} J_{n-1} \ket{n-1}\bra{n} \\
& + \sum_{n = 1}^{N} J_{n}\ket{n+1}\bra{n},
\end{split}
\end{equation}
where the couplings follow the square-root relation: $J_{n} = \frac{\pi}{2t_f}\sqrt{n\pare{N-n}}$, with $t_f$ describing the time that an initial one-site excitation takes to be transferred from site $n$ to the site $N-n+1$. It is worth mentioning that although the Hamiltonian in Eq. (\ref{Eq:PT}) was originally proposed for fermionic qubits \cite{bose2003}, its single-excitation nature suggests that it can be implemented in either quantum or classical platforms \cite{roberto-LPL}.

\begin{figure}[t!]
\centering
\includegraphics[width=8.0cm]{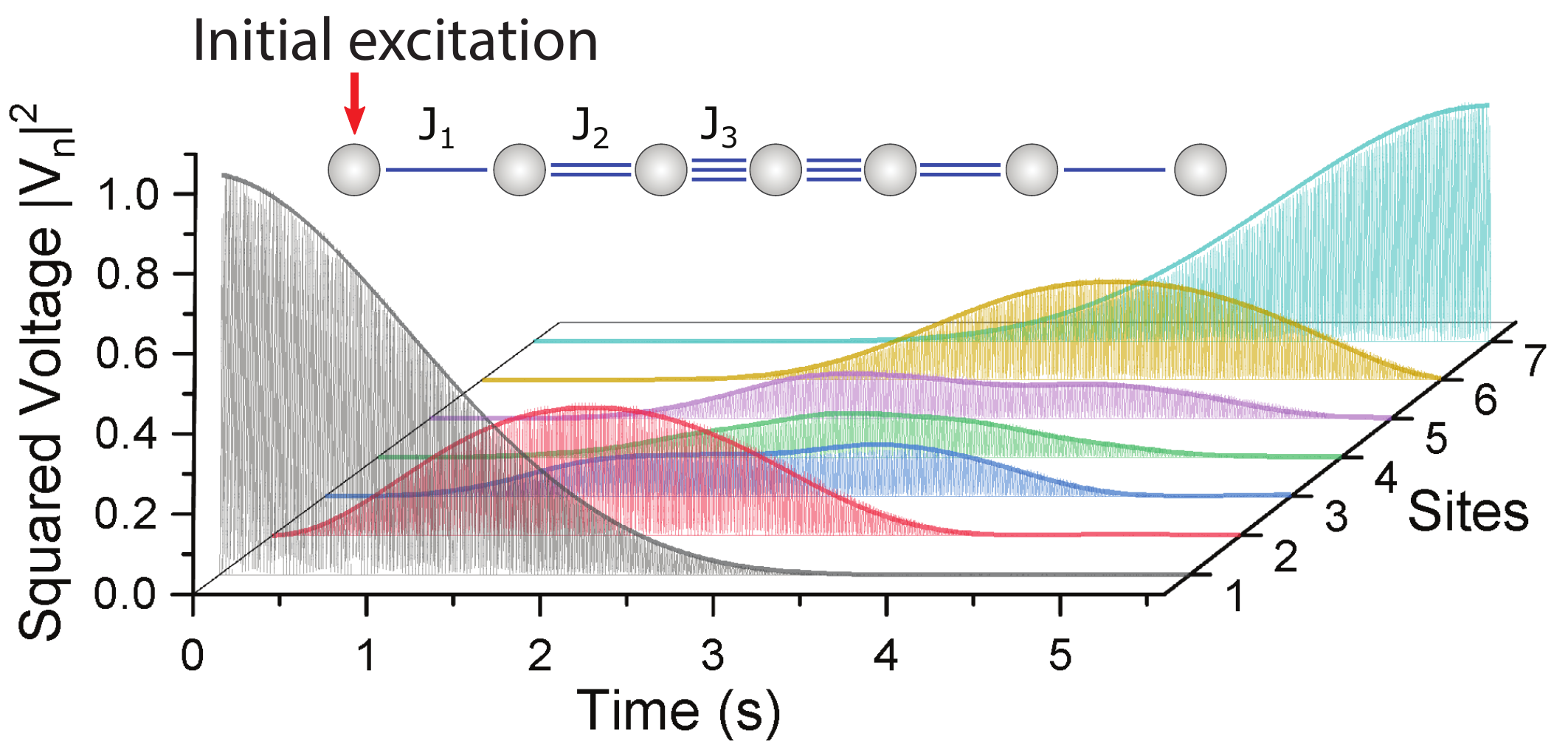}
\caption{Time evolution of a voltage signal (excitation) initialized in the first site of the chain ($n=1$). The inset shows the lattice structure, as well as the initial excitation condition. Note that after the characteristic time $t_{f}=5.6$ s, the signal injected into the first site of the chain ($n=1$) is coherently transferred to the final site (n=7) with an efficiency of 0.61, that is, 61\% of the total energy is recollected in the intended final site of the lattice.}
\label{Fig:PT_results}
\end{figure}

In an effort to provide a simple, low-cost platform for simulating quantum state-transfer protocols, we have implemented the Hamiltonian shown in Eq. (\ref{Eq:PT}). For this purpose, we have taken $N=7$ oscillators out of the ten available in the electronic platform. We have arranged them in a chain where all site frequencies, losses and couplings are characterized by the values presented in the third row of Table I in the Supplementary Materials. Note that, in order to maintain the same value for all site-frequencies, the capacitance in the oscillators take different values, this is because the corresponding frequencies strongly depend on the couplings, which change with the site positions.

Figure \ref{Fig:PT_results} shows the time evolution of an excitation (voltage signal) initialized in the first site of the chain (see the inset in Fig. \ref{Fig:PT_results}). Note that after $t_{f} = 5.6$ s, a signal injected into the first site of the chain ($n=1$) is coherently transferred to the final site ($n=7$) with an efficiency of 0.61. This means that 61\% of the total energy is recollected in the intended final site of the lattice. This rather small value is mainly due to the intrinsic losses ($>$1 k$\Omega$) of the general purpose operational amplifiers (see Supplementary Materials), which could be reduced by making use of low-noise instrumentation amplifiers \cite{OPAMP_book}.

\begin{figure}[t!]
\centering
\includegraphics[width=8.0cm]{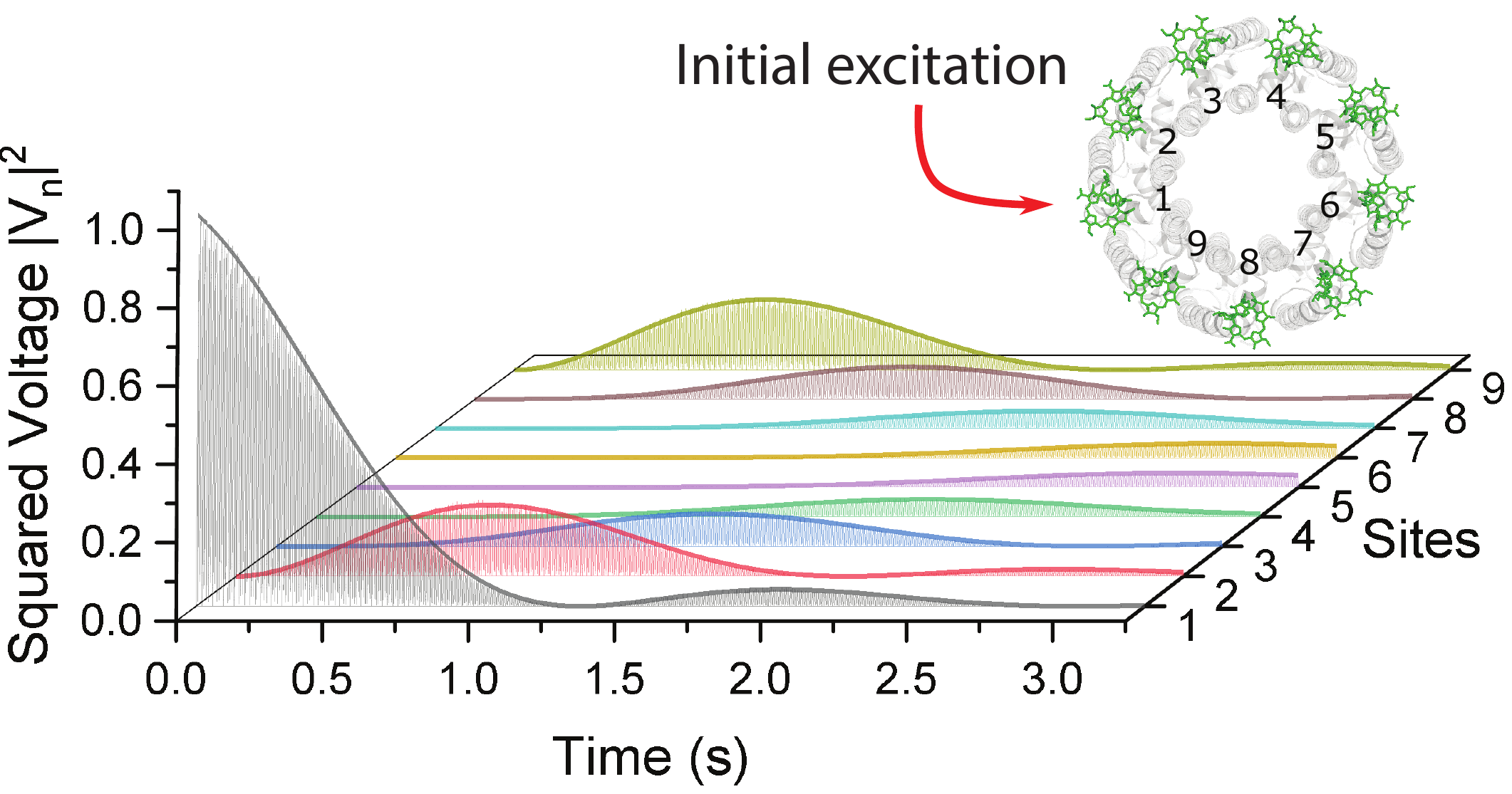}
\caption{Time evolution of a voltage signal (excitation) initialized in one of the sites of the B800 ring ($n=1$). The inset shows the ring structure, as well as the initial excitation condition. Note that the results are presented in a rescaled time-window (3.2 s), which corresponds to a $\sim 0.6$ ps time-evolution in the real photosynthetic complex.}
\label{Fig:LH2_results}
\end{figure}

\emph{Photosynthetic Energy Transport.-} We finally present, for the first time, the simulation of the exciton dynamics in the B800 ring of the purple bacteria LH2 complex. The LH2 complex of \emph{Rhodopseudomonas acidophila} carries 27 bacteriochlorophyll (BChl) molecules in two concentric rings embedded in the surrounding proteins \cite{cheng2006}, 9 of the BChl molecules form the B800 ring (see inset in Fig. \ref{Fig:LH2_results}), which absorbs maximally at 800 nm, and the other 18 molecules form the B850 ring which absorbs maximally at 850 nm. The BChl molecules in the B850 ring are closely packed, which leads to strong electronic coupling between adjacent pigments \cite{scholes_1999}, whereas the large distance between adjacent BChl molecules in the B800 ring results in a weak nearest-neighbor coupling.

In the single-excitation basis, the B800 ring can be described by a tight-binding Hamiltonian of the form \cite{cheng2006}
\begin{equation}\label{Eq:LH2}
\hat{H}_{\text{B}800} = \sum_{n=1}^{N}\pare{\varepsilon_{n}-i\Gamma_{n}}\ket{n}\bra{n} + \sum_{n \neq m}^{N} J_{nm} \ket{n}\bra{m},
\end{equation}
where the excitation energies of the BChl molecules and the coupling between them are given by $\varepsilon_{n} = 12450\;\text{cm}^{-1}$ and $J_{nm}=-27\;\text{cm}^{-1}$, respectively. To simulate the dynamics described by the Hamiltonian (\ref{Eq:LH2}), we first note that the rate at which BChl molecules interact is extremely fast compared to the characteristic frequencies of our platform. Therefore, we introduce a proper rescaling factor, which is found to be $\eta = 5.2615\times 10^{12}$. With this factor, we obtain an excitation energy of $\varepsilon_{n} = 446.4\;\text{Hz}$ and a coupling of $J_{nm}=-0.9\;\text{Hz}$. These parameters are set by making use of the values presented in the fourth row of Table I in the Supplementary Materials.

Figure \ref{Fig:LH2_results} shows the dynamics of a voltage signal (excitation) initialized in one of the sites of the B800 ring, as depicted in the inset. Note that the results are presented in a rescaled time-window (3.2 s), which corresponds to a $\sim 0.6$ ps time-evolution in the real molecular system. Moreover, note that the weak coupling between the BChl molecules in the B800 ring results in a slow propagation of the energy among the sites, thus making the system more susceptible to dissipation effects due to its interaction with an environment \cite{croce_book}.

To conclude, we have presented a versatile, reconfigurable network for the simulation of quantum transport. Our platform overcomes major limitations in existing protocols for quantum simulation, namely preservation of coherence, full control of system parameters, low losses, and scalability. We have exploited the negligible decoherence and versatility of our network to induce complex superpositions and interference effects, thus allowing us to simulate Hamiltonians attributed to important quantum transport dynamics. Because of its robustness and versatility, our device arises as a promising platform for the simulation of quantum transport phenomena.

This work was supported by CONACyT under the projects  CB-2016-01/284372 and A1-S-8317, and by DGAPA-UNAM under the project PAPIIT-IN102920. We acknowledge funding from the U.S. Department of Energy, Office of Basic Energy Sciences, Division of Materials Sciences and Engineering under Award 0000250387.

\section*{\large{Supplementary materials}}

In this document, we show how an array of $N$ inductively-coupled RLC oscillators can be electronically implemented by making use of functional blocks synthesized with operational amplifiers and passive linear electrical components. Under this scheme, we can select independently site-frequencies, couplings and losses from a broad range of possible values. Additionally, we devote a section to discuss the time-dependent energy in the quantum and
electronic models.

\section*{Circuit Design and Parameters}

\begin{figure*}[t!]
\begin{center}
\includegraphics[width=\textwidth]{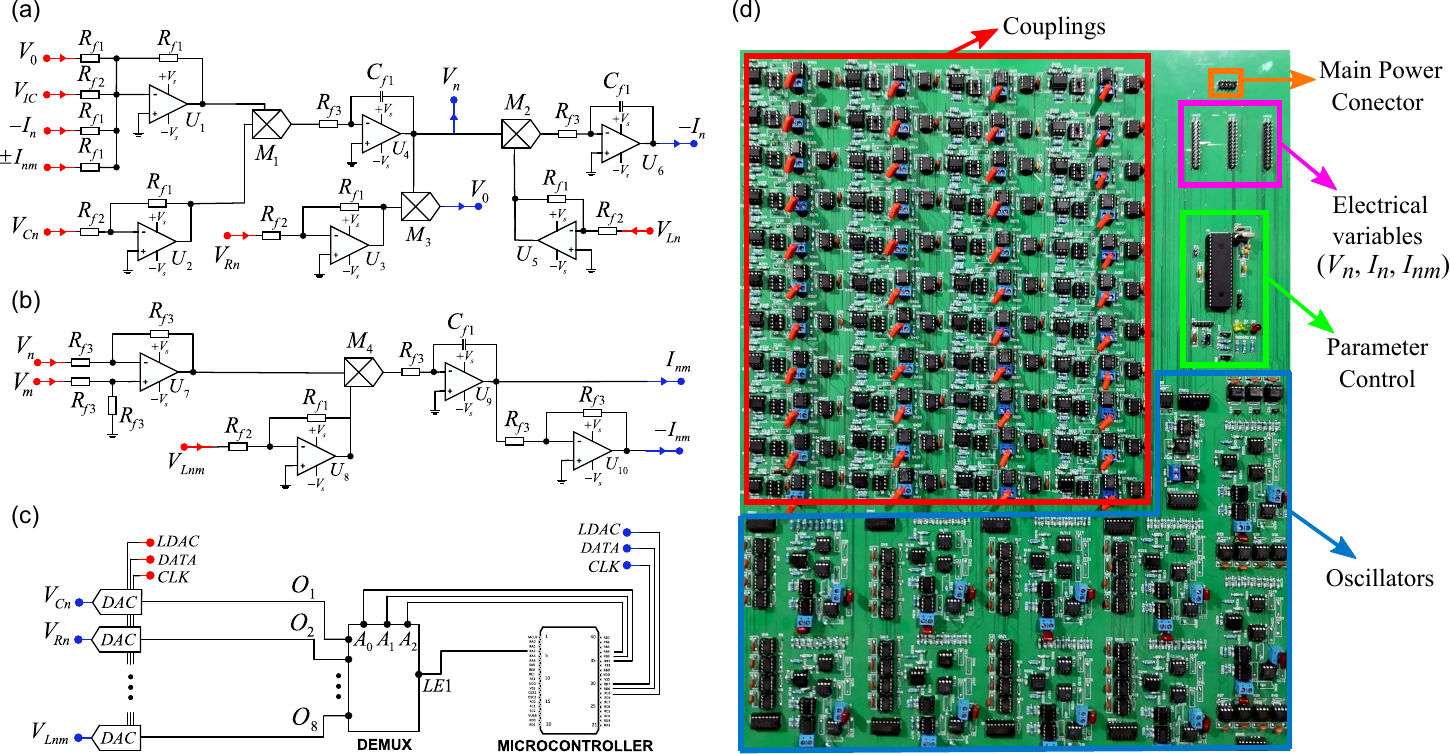}
\end{center}
\protect\caption{Schematics for (a) the oscillators, (b) the couplings and (c) the parameter digital control. The input signals are indicated with red nodes, whereas the outputs are denoted with blue ones. $R_{fj}$, $C_{fj}$, $U_j$ and $M_j$ stand for resistors, capacitors, operational amplifiers and analog multipliers, respectively. The voltage in the capacitor $V_n$, the currents across the oscillator inductor $I_n$ are the electrical variables of interest and the label $V_0$ refers to interconnection of feedback signals. $V_{Rn}$, $V_{Ln}$, $V_{Cn}$ and $V_{IC}$ are voltage signals generated by the DACs, which allow one to configure the system parameters in an easy and accessible way via software. The communication between the microcontroller and the DACs is performed through the SPI protocol, which makes use of the digital signals LDAC, DATA and CLK, and the configuration bits (LE$j$, $A_0$, $A_1$ and $A_2$) sent to the demultiplexers. (d) Printed circuit board of ten fully reconfigurable RLC oscillators.}
\label{fig:Circuit2}
\end{figure*}

Our experimental setup comprises a network of ten inductively-coupled RLC oscillators, whose dynamics are governed by Eq. (4) of the main manuscript. The oscillators and couplings are electronically implemented with active networks of operational amplifiers (OPAMPs) and passive linear electrical components. It is worth mentioning that the transfer functions of the basic electrical networks of OPAMPs namely, adders, integrators and gains, obey specific mathematical operations. This allows us to interconnect them to build complex sequences of mathematical functions where voltage signals represent physical variables of the system that is being studied. In this representation, the parameters of the system are mapped into passive components within the active networks, such as resistors and capacitors. Consequently, any change in the parameters leads to physically replacing components.

To avoid this, we have merged basic electrical networks with integrated analog multipliers for synthesizing voltage-driven components whose values depend on an external voltage signal provided by digital-to-analog converters (DAC) that communicate with a master microcontroller by the serial peripheral interface (SPI) protocol. Because of this remarkable feature, the initial conditions and the system parameters $R_n$, $L_n$, $C_n$ and $L_{nm}$, which control the site-frequencies and couplings, can be individually addressed within a wide range values via software. More importantly, since coupling values can be set to zero, one can control the connection topology between oscillators by enabling or disabling the couplings.

Structurally, our experimental setup is divided in two parts, analog and digital. The former encompasses the oscillators and couplings, both of them built with purely analog electronic components. Figures \ref{fig:Circuit2}(a) and (b) show the general schemes of the electronic circuits for the oscillators and couplings, respectively. There, $R_{fj}$, $C_{fj}$, $U_j$ and $M_j$ stand for metal resistors (1\% tolerance), polyester capacitors, general-purpose operational amplifiers LF353 and analog multipliers AD633JN (four-quadrant voltage multiplier), respectively.

In Figure \ref{fig:Circuit2}, the input signals are indicated with red nodes, whereas the outputs are denoted with blue ones. $V_n$, $I_n$ and $I_{nm}$ are the electrical variables of interest, namely the voltage in the capacitor, the currents across the oscillator inductors and the coupling inductors, respectively. The label $V_0$ refers to interconnection of an internal signal. In the experimental setup, the parameters of the oscillators and couplings, $R_n$, $L_n$, $C_n$ and $L_{nm}$, as well as the initial conditions $V_n(0)$ are defined by the values of $R_{fj}$, $C_{fj}$, $V_{Rn}$, $V_{Ln}$ $V_{Cn}$ and $V_{IC}$. These quantities satisfy the following relationships

\begin{eqnarray}
\frac{1}{R_n}&=&\frac{R_{f1}V_{Rn}\phi}{R_{f2}}, \nonumber \\
\frac{1}{L_n}&=&\frac{R_{f1}V_{Ln}\phi}{R_{f2}R_{f3}C_{f1}}, \nonumber  \\
\frac{1}{C_n}&=&\frac{R_{f1}V_{Cn}\phi}{R_{f2}R_{f3}C_{f1}},
\\
\frac{1}{L_{nm}}&=&\frac{R_{f1}V_{L{nm}}\phi}{R_{f2}R_{f3}C_{f1}}, \nonumber \\
V_n(0)&=&\frac{R_{f1}V_{IC}}{R_{f2}}, \nonumber
\end{eqnarray}
where $\phi=1/10$ is a manufacturing default factor of the analog multiplier, integrated to avoid saturation of the output voltage. Remarkably, the $R_{fj}$ and $C_{fj}$ devices represent the core configuration of the electronic platform and fix the maximum values that the system parameters can take. Furthermore, the voltage signals $V_{Rn}$, $V_{Ln}$, $V_{Cn}$ and $V_{IC}$, coming from the DACs, and taking discrete values between 0 V and 5 V with a resolution of 1.22 mV, allow to independently select such parameters from a broad range of possible values within the defined interval. To operate the operational amplifiers and analog multipliers in a convenient bandwidth, the resistor and capacitor values are set to $R_{f1}=10$ $\text{k}\Omega$, $R_{f2}=5$ $\text{k}\Omega$, $R_{f3}=1$ $\text{k}\Omega$ and $C_{f1}=0.1$ $\mu$F. This configuration allows us to tune the site frequencies from $0$ Hz to $1590$ Hz. Finally, to energize the device, we make use of a stabilized DC power supply (KEITHLEY triple channel, 2231A-30-3), which feeds the $\pm$12 V bias voltage $(+V_s,-V_s)$ to the OPAMPs and the analog multipliers.

\begin{table*}[t!]
\setlength{\arrayrulewidth}{0.2mm}
\setlength{\tabcolsep}{0.5mm}
\setlength{\doublerulesep}{0.6mm}
\extrarowheight = -0.5ex
\renewcommand{\arraystretch}{1.7}
\begin{tabular}{c c c}
\arrayrulecolor{black} \hline
\rowcolor[HTML]{aad5ff}
\textbf{Experiment} & \makebox[5.5cm][c]{\textbf{Site-Frequencies and Losses}} & \textbf{Coupling Coefficients} \\
 \rowcolor[HTML]{f0f8ff}
                       &  $C_n = 1.50$ mF            &   \makebox[5.3cm][c]{$L_{nm} = \cor{L_{x}\pare{1 - \Delta}, L_{x}\pare{1 + \Delta}}$ }  \\ \rowcolor[HTML]{f0f8ff}
 Anderson Localization &   $L_n = 3.35$ mH            &  $L_x = 96.05$ mH     \\ \rowcolor[HTML]{f0f8ff}
                       &   $R_n = 1\;\text{k}\Omega$         &  $\Delta = 0,0.5,0.9$     \\
\arrayrulecolor{white}\hline \hline \rowcolor[HTML]{f0f8ff}
                & $C_n = 1.50$ mF      &  $L_{\alpha} = 96.05$ mH \\ \rowcolor[HTML]{f0f8ff}
The SSH Model & $L_n = 3.35$ mH      &  $L_{\beta} = 192.1$ mH \\ \rowcolor[HTML]{f0f8ff}
                & $R_n = 900\;\Omega$  &  \\
\hline \hline  \rowcolor[HTML]{f0f8ff}
                   & $C_1 = C_4 = C_7 = 7.54$ mF & $L_{12}=L_{67}=321.36$ mH \\ \rowcolor[HTML]{f0f8ff}
\makebox[4.6cm][c]{Coherent Transfer of States}  & $C_2=C_3=C_5=C_6=7.58$ mF   & $L_{23}=L_{56}=181.97$ mH     \\ \rowcolor[HTML]{f0f8ff}
                   & $L_n = 1.11$ mH             & $L_{34}=L_{45}=75.45$ mH    \\ \rowcolor[HTML]{f0f8ff}
                   & $R_n = 1.5\;\text{k}\Omega$            &       \\
\hline \hline  \rowcolor[HTML]{f0f8ff}
                          & $C_n = 1.50$ mF           &     \\ \rowcolor[HTML]{f0f8ff}
Photosynthetic Transport  & $L_n = 3.35$ mH           &   $L_{nm} = 806.90$ mH   \\ \rowcolor[HTML]{f0f8ff}
                          & $R_n = 1\;\text{k}\Omega$ &     \\ \arrayrulecolor{black} \hline
\end{tabular}
\caption{Electrical-component values used in the implementation of the quantum transport protocols presented in the main article.}
\label{Tab:table1}
\end{table*}

As for the digital part, we incorporate digital-to-analog converters (MCP4921, Resolution 12 bits), demultiplexers (SN47HC138N, high speed CMOS 3-to-8 line decoder) and a microcontroller (PIC18 familiy), which together deal with the parameter and initial condition configurations. Note that in the electronic platform there are eighty-five configurable parameters, four per each oscillator and forty possible all-to-all couplings, each one of them controlled by voltage signals coming from the DAC ($V_{Rn}$, $V_{Ln}$, $V_{Cn}$ and $V_{IC}$). To satisfy this demand, the enable/disable terminal of each DAC is connected to a digital bus managed by demultiplexers, in this way, with only sixteen lines of the microcontroller we can select a particular DAC, setting the properly configuration bits, $LEj$, $A_0$, $A_1$ and $A_2$, to the demultiplexers, as well as to transmit a desired output voltage to the DAC through the SPI protocol using the control and data bits, namely, LDAC, DATA and CLK [see Fig. \ref{fig:Circuit2}(c)].

To ensure a strong connection among the electronic components, we design and manufacture a printed circuit board (PCB, 40$\times$50 cm) where electronic devices were mounted and soldered. The PCB was designed in Altium Software and fabricated with a computer numerical control (CNC) laser. The electronic realization of the ten fully reconfigurable RLC oscillators on the PCB is shown in Fig. \ref{fig:Circuit2}(d). The module controlling the parameters and initial conditions is indicated with a green square, whereas the analog oscillators and couplings are signaled with blue and red, respectively. Both the analog and digital modules of our experimental setup are energized through the main power connector (orange squared). The acquisition of the electrical variables (magenta square) is performed with a Digilent oscilloscope (Analog discovery 2), which directly transfers the information to a computer by USB connection.

To conclude this section, we finally provide (in Table \ref{Tab:table1}) detailed information regarding the electronic-component values needed for the implementation of the experiments described in the main text.

\section*{Energy Loss Estimation}

In this section we provide a thorough description of how the unavoidable losses, present in our electronic platform, can be accounted for in the quantum tight-binding network model. Let us consider the energy contained in the whole circuit, which is given by
\begin{equation}
Q_{\text{cl}}(t)=\frac{1}{2}\sum_{m \neq n} C_nV_n^2+L_nI_n^2+L_{nm}I_{nm}^2.
\end{equation}
As one might expect, in the presence of losses (or resistance), the total energy of the system will decay following an exponential behavior \cite{roberto2018}. Of course, in the absence of resistance, the total energy is conserved. Remarkably, in the quantum model, a quantity that follows the same behavior in the presence (or absence) of losses is the trace of the density matrix. We define as an energy-like measure of the quantum system, given by the expression
\begin{equation}
Q_{\text{q}}(t)=\sum_{n}\left|c_n\right|^2.
\end{equation}
Indeed, it is well known that for a closed quantum system, the trace of the system's density matrix is conserved, whereas for a system affected by a dissipative environment, the trace of the reduced density matrix (defined as the density matrix obtained after the environment's degrees of freedom are traced out) decays exponentially \cite{opensys_book}. Along this line, the simplest way of introducing a dissipative process in a quantum system is by including a non-Hermitian term, in the closed system's Hamiltonian, of the form
\begin{equation}
\hat{H}_{\text{loss}} = -\frac{i}{2}\sum_{n}\Gamma_{n}\ket{n}\bra{n},
\end{equation}
with $\Gamma_{n}$ describing the rate at which energy is dissipated to the system's environment.

\begin{figure}[t!]
\centering
\includegraphics[width=8.25cm]{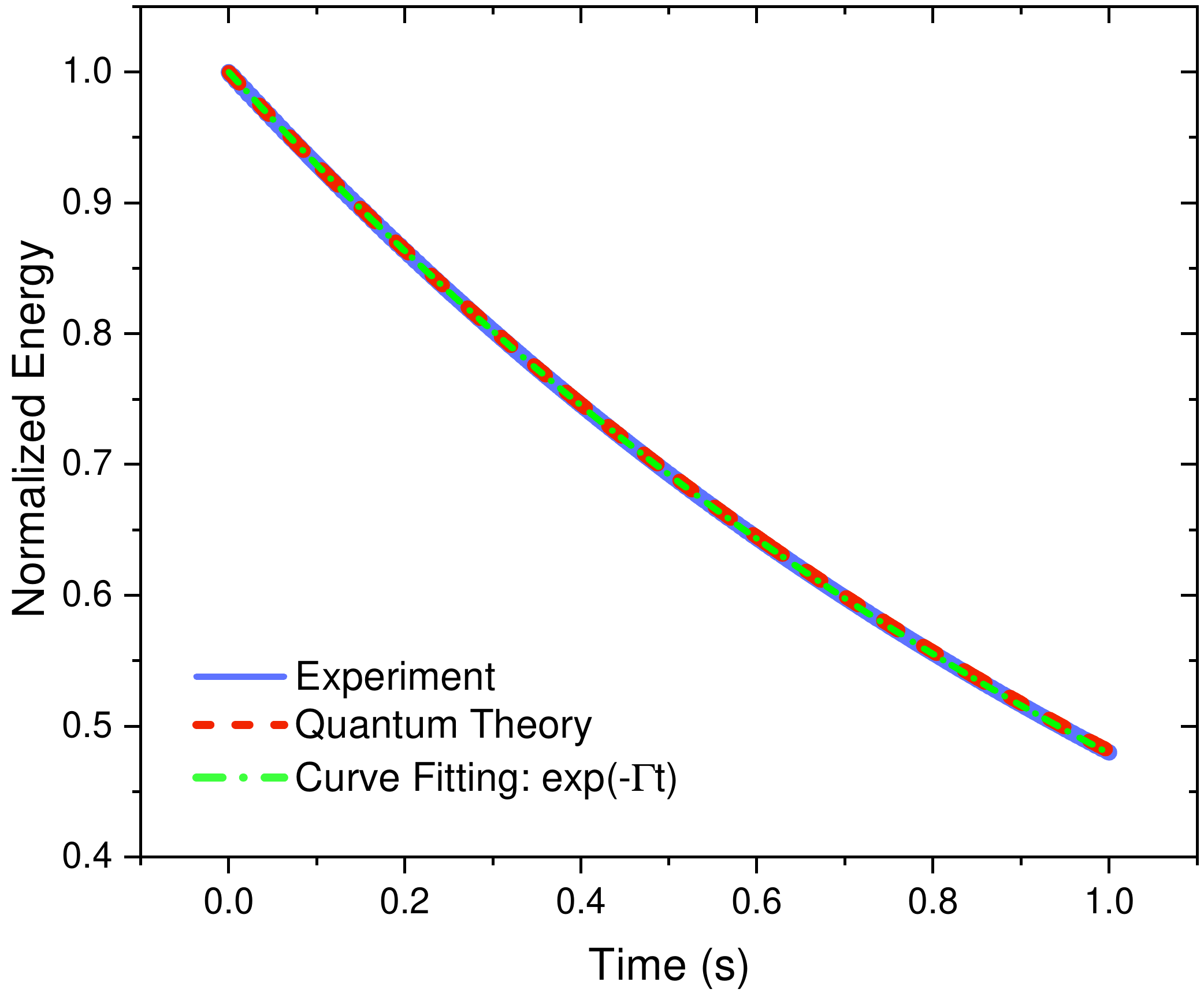}
\caption{Time evolution of the normalized energy in the electronic platform (solid line) and the quantum dissipative model (dashed line). The dash-dotted line shows an exponential curve fitting. The parameters used for obtaining the energy-curves are those used in the experimental implementation of the SSH model.}
\label{Fig:Energy}
\end{figure}

Figure \ref{Fig:Energy} shows the time evolution of the normalized energy for the electrical circuit (blue solid line) and the trace of the open system's density matrix (red dashed line). Note that both curves follow the same exponentially-decaying behavior (described by the green dotted curve fitting), which allows us to establish the relation: $\Gamma = 1/\pare{RC}$, where $R=R_n$ and $C=C_n$ stand for the resistance and capacitance of each electrical oscillator in the device, respectively. This important result is what allows us to include the effects of the electronic parasitic losses into the quantum tight-binding model.

\providecommand{\noopsort}[1]{}\providecommand{\singleletter}[1]{#1}%

\end{document}